\documentclass[useAMS,usenatbib]{mn2e}
\usepackage{epsfig}
\usepackage{natbib}
\usepackage{color}
\usepackage{soul}
 \usepackage{aas_macros}
\title[ Daily Ceres Albedo changes ]{ Daily   variability of  Ceres' Albedo  detected by means of radial  velocities   changes of the reflected sunlight}
\author[]{ P. Molaro$^{1}$, A.~F.~Lanza$^{2}$, L. Monaco $^{3}$, F. Tosi, $^{4}$, G. Lo Curto $^{5}$,
 M. Fulle$^{1}$, L. Pasquini $^{5}$\thanks{E-mail:
molaro@inaf.oats.it (PM)}, 
 \footnotemark[1]\thanks{ Based on observations collected at the European Souther Observatory, Chile.  Program
 ESO  DDT 295.C-5031, 5035.
 }\\
$^{1}$  INAF-Osservatorio Astronomico di Trieste, Via G.B. Tiepolo 11, I-34143 Trieste, Italy\\
 $^{2}$  INAF-Osservatorio Astrofisico di Catania, Via S.~Sofia, 78, 95123 Catania, Italy \\
  $^{3}$ Departamento de Ciencias Fisicas, Universidad Andres Bello, Repuœblica 220,  Santiago, Chile\\
 $^{4}$ INAF-IAPS INAF-IAPS Istituto di Astrofisica e Planetologia Spaziali, Via del Fosso del Cavaliere, 100, I-00133 Rome, Italy\\
  $^{5}$ ESO, Karl-Schwarzschild-Strasse 2, D-85748 Garching bei Munchen, Germany
 }
\begin{document}

\date{Accepted.... Received 2012}

\pagerange{\pageref{firstpage}--\pageref{lastpage}} \pubyear{2002}

\maketitle

\label{firstpage}

\begin{abstract}

Bright features  have been   recently discovered  by Dawn on Ceres, which     extend previous   photometric and Space Telescope observations.  These features should produce distortions  of the   line profiles of the reflected solar spectrum    and therefore an apparent radial velocity variation modulated by the rotation of the dwarf planet.   Here we report on two sequences of  observations of    Ceres  performed in the nights of 31 July,  26-27  August 2015 by means of the high-precision HARPS spectrograph at the 3.6-m La Silla ESO telescope.  The observations revealed a quite complex  behaviour which likely combines a radial velocity modulation due to the rotation   with an  amplitude of  $\approx \pm 6$~m~s$^{-1}$  and  an unexpected   diurnal effect.
The latter changes   imply  changes in the albedo of   Occator's  bright features 
 due to the blaze produced by the exposure to    solar radiation.   
The    short-term variability of Ceres' albedo is on  timescales  ranging from hours to months and can both be confirmed and followed by means of dedicated radial velocity  observations.

\end{abstract}

\begin{keywords}
Planets and satellites: general   - -- Stars: planetary systems --  
\end{keywords}

\section{Introduction}
 Ceres'    diameter of 950~km  makes it  the largest  body in the asteroid belt as well as  the smallest  dwarf planet in the Solar System. At least one-quarter of its  mass is composed of water, a proportion greater than   in other asteroids  and  even    on Earth.

 \citet{cha07}    obtained  Ceres' light curves   from 1958 to 2004  and found  an optical flux modulation of $\sim 0.045$~mag  which  provided a  very precise rotational period of 9.074170 $\pm$ 0.000002  h.  
 In 2003 the Hubble Space Telescope  captured   a   spot  moving with  Ceres'  rotation.    Hubble images   acquired  by the Advanced Camera for Surveys  at a resolution of 30 km in three wide band filters (535, 335 and 223 nm) allowed the identification of eleven surface albedo features ranging in scale from 40 to 350 km  but with no details revealing  their origin \citep{li06}.   Water vapour plumes erupting off the surface of Ceres have been observed with  Herschel  by \cite{kup14} who suggested that they  may be produced by volcano-like ice geysers. 
NASA's Dawn Spacecraft    reached   its  final destination to Ceres in 2015 and the first    images of  the asteroid   revealed  the  unexpected presence of  a prominent  spot in its  northern hemisphere in coincidence with feature N.5 of \citet{li06}, which was   mainly seen in the UV bands and barely in the optical.  In the images taken in February by Dawn at a distance of 46000 km one  of 
Ceres' bright spots located in the Occator crater  revealed  a companion of lower brightness, but apparently in the same basin.  Images taken in June  showed that the bright spot  consists of a large bright area at its center and several smaller spots nearby. The presence of several  bright spots   in the same basin  may be pointing  to a volcano-like origin of the spots. However, it is  not  clear how   an isolated  dwarf planet could be thermodynamically active enough to generate either of them.  
 
It has been suggested that the bright  spots  may be located in  a relatively fresh giant  impact crater  revealing  bright  water ice under a thin crust.       It has also been  speculated that      they could be evidence of cryo-volcanism or even icy geysers \citep{wit15,kup14}.  Images by Dawn  reveal the spots even when they are near the limb of Ceres, when the sides of the impact crater would normally block the view of anything  placed at the  bottom.  This   suggests that the  main feature  rises relatively high above the surface of the crater.
 Dawn scientists have not yet established   whether the bright spots are made of ice, of evaporated salts, or something else.  Thus up to this day the nature of these  spots   remains a mystery. 

Small rocky bodies  are  reflecting  almost unaltered sunlight as point-like sources,  besides being also good radial velocity standards    \citep{mol12}.  However, the presence of an albedo inhomogeneity    implying  either an enhanced  or reduced  reflectivity in a small area of the rotating body  should produce a distortion in the reflected solar line profiles modulated by the asteroid rotation \citep{lan15b}. 
Therefore,  Ceres photometric variability   should be associated with  a modulation of the radial velocity of the reflected solar spectrum with the rotational  period at the level of few  m s$^{-1}$. 
This phenomenon is similar but opposite  to the  Rossiter Mc-Laughlin effect occurring  from the {transit  of a  planet  in front of the  host star}  \citep{que00, mol13}.  
   
\section{Observations and kinematics}
 We observed Ceres on 30 July 2015 taking  a sequence of  40 HARPS exposures of  780 sec  each. The observations started  on   30 July at  23:33 UT and ended  on  31 Jul  at 09:14 UT (MJD= 2457233.994 to 2457234.381) in an unbroken sequence lasting  for 9.45 hours  which is slightly longer than Ceres rotational period of 9.074 hours. 
On the 26th of August we  took a second sequence of 29  exposures of 900 sec each. The observations started  on 26 August at 23:36 UT at the beginning of the night and  ended at the set of Ceres on  27 Aug at 06:49 UT  (MJD 2457260.9878- 2457261.2840) in an uninterrupted sequence  lasting 7.25 hours, i.e. a bit shorter than a complete rotation.    Five exposures  were also taken  at the end of the night on 27th August  when we could shortly open the dome during a  pause of  bad  weather. The journal of the observations together with the relevant quantities are  given in Tab \ref{table:1}.
 
 Ceres had    an  APmag of $\approx$ 7.5 mag, an angular dimension  of 0.69 arcsec and was observable during  most of the night.  The observations  were performed  at  phase angles of 4.4  and 12.3 degrees in July and August,  respectively.  At opposition   Ceres becomes slightly brighter    due to the Opposition Surge effect,    but  no radial velocity variations are expected within a single rotational period \citep{mol15b}. Inspection of the solar activity through the Solar Dynamic Observatory revealed no significant change in the solar activity to induce an important radial velocity change. The solar activity measured from the  total  sunspot area changed slightly from  $\approx$ 500 to $\approx$ 1250 millionths of solar hemisphere between  31 July  and 27 August. However, note that these will introduce a radial velocity  offset between the two nights but  not a change within one epoch. 

 We used  HARPS    pipeline to obtain the radial velocities from sunlight spectra which are reported in the third column of  Tab \ref{table:1}.  In the HARPS spectrograph 
 a second fiber  supplies   simultaneous ThAr spectra that  are  used to correct for
instrumental radial velocity drifts occurring over the night due to residual temperature, pressure or mechanical changes. The radial 
velocity differences
with respect to the previous calibration provide the instrumental drifts
of the spectrograph.  We are considering the  values RV$_c$  corrected for instrumental drift and reported in the   column 4 of online Tab \ref{table:1} but note that corrections are always below 0.5~m~s$^{-1}$ and  do not affect  the   results.
Single observations have S/N $\ge$ 100 and  the  error in the radial velocity measurement is lower than 1 ~m~s$^{-1}$.
The  pipeline radial velocity RV$_c$ is relative to the solar system 
barycenter, but it is not  appropriate for reflecting bodies. Thus,  we  remove   the pipeline barycentric radial velocity correction, the {\it 
BERV},     and    the proper  radial velocity   becomes: 
 \begin{equation}
\centerline{$ RV = RV_c -  BERV   -   ( \dot r + \dot \delta ).$}	
 \end{equation}
 
where   the radial velocity of the
observer  relative to  Ceres  is  $\dot \delta $   and   $\dot r$ is that   of Ceres relative to the Sun. The latter   needs  to be also considered   
    because the sunlight reflected by  Ceres   is shifted by its  
radial velocity   with respect to the Sun at the time   the photons 
leave  Ceres  \citep{mol11, lan15b}.   The quantities  $\dot r $ and $ \dot \delta$ are  computed  by
using the JPL horizon ephemerides \footnote{Solar System Dynamics Group, 
Horizons Web Ephemerides Systems, JPL, Pasadena, CA 91109 USA 
http://ssd.jpl.nasa.gov}.
%The component of the   observer on   Earth  
%  is  the projection of the asteroid motion along the line of sight 
 % adjusted for aberration, and includes both 
%the radial velocity of  Ceres   and the components  from the Earth's  rotation and revolution {\bf (\dot \delta)}.. 
%We do not correct for the relativistic components which  amount to a few m~s$^{-1}$  since they do not vary along the night   and are     incorporated into  the zero baseline.
%The change between the two epochs is of 0.204 m~s$^{-1}$ and can be neglected.}
The relativistic components are of  0.766  m~s$^{-1}$  in July and of 0.957 m~s$^{-1}$  in August with 
a  change between the two epochs  of $\approx$ 0.2  m~s$^{-1}$ which can be neglected.
  The error in these estimations depends on the precision of the ephemerides but should be  of few ~cm~s$^{-1}$.
 The average rate  in the radial velocity change due to kinematical motions alone   is 
$\approx$ 0.65 -0.67~m~s$^{-1}$    per minute.  During the 
exposure of 15  minutes this velocity
  change produces a small  but finite spectral smearing. However, the spectral smearing is symmetrical  to a  
good approximation  and  it does not result into  a net shift in the  radial 
velocities estimated for  flux weighted    mid-exposure values.
 
 The   HARPS pipeline   cross--correlates the  reflected solar spectrum  with
   the Fourier Transform Spectrometer 
(FTS) solar spectrum obtained by Kurucz at the McMath-Pierce Solar Telescope at Kitt Peak 
National Observatory \citep{kur84}.  The FTS solar spectrum  is  anchored in wavelength onto 
few  telluric emission lines and  has an uncertainty associated in the  zero point of the order of 100~m~s$^{-1}$
\citep{kur84,mol12}. 
   This offset  was  measured 
   in coincidence of the Venus transit  of 6 June 2012 and  was found to be of  102.5~m~s$^{-1}$ \citep{mol13} with an   
 uncertainty of the order of few~m~s$^{-1}$. Note that this value also incorporates    the   solar 
activity  on that day. 
However,  in June 2015  a new set of octagonal fibers have replaced the original circular ones as part of the HARPS upgrade. 
This led to a change in the Instrumental Profile of the spectrograph and  a shift in the RV-offset. Measurements of a set of RV standards before and after the upgrade  show that the shift  depends on the   line shape and on the stellar spectral type. For two G2V solar twins the shift has been found  increased by +16.1 and  14.1     m s$^{-1}$  if compared to previous values obtained with the circular fiber, i.e. about 118  m s$^{-1}$ in total  \citep{loc15}.

\begin{figure*}
\centering
 \includegraphics[width=17cm]{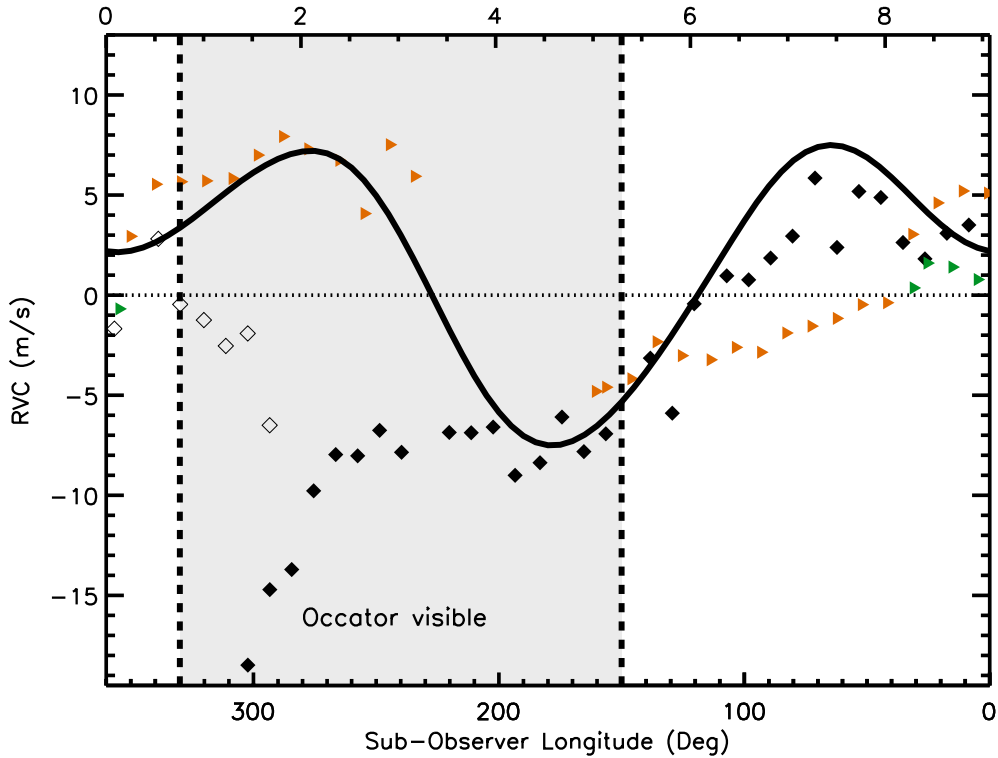}
 \caption{  Radial velocities versus the East sub-observer longitude in the reference frame. The longitude decreases with time (indicated in hours on the top axis) or the rotation phase.  Black  diamonds are the measurements taken in  July, filled diamonds are for  the beginning of the night and empty diamonds  for  the end.  Orange triangles are the observations of  26  August and green  triangles of the  27 August. A baseline   of 115.4~m~s$^{-1}$ is subtracted in the observations. 
 %The July sequence starts with the value at longitude $\approx$ 300$^0$ and RV$_c$ $\approx$ 98  m~s$^{-1}$ and rises monotonically.   The  observations in August start at phase longitude of 150$^0$ deg. 
 The symbol size is comparable to the expected radial velocity error, i.e. at sub ~m~s$^{-1}$ level.
   The  black continuous line shows the model  of RV changes based on  Chamberlain et al.'s photometric template light curve (see the text). The shadowed area is when Occator is visible. 
%   The model is shifted vertically by 116.5~m~s$^{-1}$,  which is about the expected  offset \citep{loc15}, and scaled in amplitude to reproduce the data. 
 %  The effect of the two light curve maxima are also shown  individually with red dotted and cyan dashed lines.  To make them clearly visible, they are shifted vertically by   -0.5~m~s$^{-1}$. 
%   The dashed vertical lines mark the region of visibility  of  Occator.
}
 \label{fig1.pdf}
\end{figure*}
 
 The results are shown in  Fig. 1,  where the  radial velocity is plotted vs. the sub-observer longitude  in the reference frame adopted by  \citet{cha07}. The longitude increases  in the direction of the dwarf planet  rotation, so that the longitude of the sub-observer point decreases vs.  time.   The   diamonds refer to the measurements taken in  July  while the orange and green triangles are the August observations.   Note that the July sequence starts at longitude $\approx$ 300$^{\circ}$. The higher values at the same  phase  are at the end of that night and are shown with empty diamonds.  The peak-to-peak variability in the radial velocities  is of about 15 m s$^{-1}$ while the photon   radial velocity error is of $\approx$ 1  ~m~s$^{-1}$  and  no evidence for larger systematic errors, which, however, could always plague in.
 To be noted that \citet{lan15}  traced the  radial velocities of Ceres and other bodies in the solar system since 2006  and  observed an excess in the  scatter in the radial velocities of Ceres with respect to the other bodies. In 12 years of  temporal span the standard deviation around the mean is of 6.6  m~s$^{-1}$ , consistent with the dispersion observed here, while for the others bodies studied is in the range of 2 to 4 m~s$^{-1}$  revealing a peculiar behaviour of Ceres's radial velocities.
 
The data points are folded with the rotational phase and is possible to see that the observations do not overlap between two successive rotational periods
  and also between the  measurements taken about one month apart. The   observed radial velocity pattern    which does not reproduce over   two close rotational periods  necessarily  implies a certain degree of variability of the main albedo features responsible for the radial velocity variations    on a very short temporal scale.

 \begin{table*}
\caption{Journal of July observations. The full table is available online.}             
\label{table:1}      
\centering          
\begin{tabular}{c c c  cc    cc   c c c  }     % 10 columns 
\hline\hline       
Date & MJD  & RV &   RV$_c$ & BERV  & Exp & T'  & JDMIDEXP & $\dot r$ & $\dot \delta$\\ 
 &    & m~s$^{-1}$  &    m~s$^{-1}$  &  m~s$^{-1}$    & s &   &  m~s$^{-1}$   &  m~s$^{-1}$   &   m~s$^{-1}$  \\ 
\hline  
2015-07-30 23:33:06 &  2457233.98138201 & 914.885 &	 914.578    &  -3203.246    &	   179.9  &	 0.50  &   2457234.48131   & 671.421  & 3357.268 \\
2015-07-30 23:51:29 &  2457233.99414785 & 917.027 &	 916.741    &  -3224.302    &	   779.9  &	 0.54  &   2457234.49408   & 671.371  & 3371.657 \\
2015-07-31 00:11:23 &  2457234.00353524 & 926.589 &	 926.437    &  -3236.302    &	   779.9  &	 0.49  &   2457234.50790   & 671.317  & 3389.635 \\
2015-07-31 00:25:32 &  2457234.01292216 & 927.485 &	 927.276    &  -3250.686    &	   779.9  &	 0.54  &   2457234.51773   & 671.279  & 3403.891 \\
2015-07-31 00:38:32 &  2457234.02230851 & 931.640 &	 931.080    &  -3264.919    &	   779.9  &	 0.50  &   2457234.52675   & 671.244  & 3418.031 \\
%2015-07-31 00:52:08 &  2457234.03175053 & 932.855 &	 932.761    &  -3280.839    &	   779.9  &	 0.50  &   2457234.53620   & 671.207  & 3433.858 \\
 \hline      	    																						 
 \end{tabular}	    																						 
\end{table*}

  \section{A model}
The light curves of most asteroids are related to their irregular shapes.  However,  Ceres' smooth oblate
spheroidal shape suggests that  the light curve variations observed by \citet{cha07}  come from albedo features.   
   These  light curves   show variations in the V band of    0.04-0.06 
mag, i.e.  about  3.6-5.4 percent, depending on the phase angle. 
%They provide a template light curve giving   $\Delta V$ vs. the sub-Earth longitude $\ell$ in their reference frame.  
There are  two photometric maxima  at phases of about  $0^{\circ}$ and $110^{\circ}$.
The  one  at phase  $110^{\circ}$ is the largest with  a flux variation of approximately 3.1 percent compared with 2 percent of the secondary one.   
Moreover, there are also two photometric minima at  phases of about  $30^{\circ}$ and $240^{\circ}$.
 
We used  this  template to   model the RV modulation induced by the  inhomogeneities of the albedo. We assume that Ceres has a spherical shape  and subdivide its surface  into $N=7200$  elements of size $3^{\circ} \times 3^{\circ}$. The colatitude of the $k$-th element measured from the North pole and its longitude are indicated  as $(\theta_{k}, \ell_{k})$.  The albedo of the surface elements is assumed to be uniform, except for those falling within a belt of $\pm 15^{\circ}$ across the equator that is varied to reproduce the observed template light curve.  Specifically,  we computed the relative flux variation $\Delta f /f_{0}$ observed at sub-Earth longitude $\ell$, where $f_{0}$ is the reference flux that corresponds to  $\Delta V_{\rm mean}$, i.e.  the mean value of $\Delta V$:
\begin{equation}
\frac{\Delta f(\ell)}{f_{0}} = 10.0^{-0.4[\Delta V(\ell) - \Delta V_{\rm mean}]} -1
\end{equation}
We  assume that the flux variation comes from an albedo feature at longitude $\ell$: $a(\ell) = a_{0} [1 + c_{\rm f} \times (\Delta f(\ell)/ f_{0})]$, where $a_{0}$ is the mean albedo and $c_{\rm f}$ a factor that is adjusted to reproduce the amplitude of the template light curve because we have no information on their  latitudinal extension.
To compute the radial velocity  at each rotation phase we compute the cosine of the angles $\psi_{k \odot}$ and $\psi_{k \oplus}$ between the normal to the $k$-th surface element and the Sun-asteroid or the asteroid-observer directions, respectively  \citep{lan15b}:
\begin{eqnarray}
\mu_{k\odot} = \cos \psi_{k \odot} & = & \sin i_{\odot} \sin \theta_{k} \cos [\ell_{k} + \Omega (t-t_{0})+\Delta \phi] \nonumber \\
& & + \cos i_{\odot} \cos \theta_{k},  \\
\mu_{k\oplus} = \cos \psi_{k \oplus} & = & \sin i_{\oplus} \sin \theta_{k} \cos [\ell_{k} + \Omega (t-t_{0})] \nonumber \\ 
& & + \cos i_{\odot} \cos \theta_{k},  
\end{eqnarray}
where $i_{\odot}$ and $i_{\oplus}$ are the inclination angles of the Sun-Ceres or the Earth-Ceres lines to the spin axis of the asteroid, respectively, $\Omega = 2\pi/P_{\rm rot}$ its angular velocity of rotation  with the rotation period $P_{\rm rot} = 9.074170$ hours, $t$ the time, $t_{0}$ the initial time, and $\Delta \phi$ the phase angle. The spin axis of Ceres is almost perpendicular to the plane of the ecliptic, therefore we assume $i_{\odot} = i_{\oplus}=90^{\circ}$ and consider a mean value for the phase angle of $\Delta \phi = 8^{\circ}.35$ for our time period. The radial velocity of the $k$-th surface element towards the Sun or the observer produced by Ceres' rotation at the time $t$ is given by, respectively:
\begin{eqnarray}
\label{vr_eqs1}
v_{k \odot} &  = & -V_{\rm eq} \sin i_{\odot} \sin \theta_{k} \sin   [\ell_{k} + \Omega (t-t_{0})+\Delta \phi] \\
v_{k \oplus} & =  & -V_{\rm eq} \sin i_{\oplus} \sin \theta_{k} \sin   [\ell_{k} + \Omega (t-t_{0})], 
\label{vr_eqs}
\end{eqnarray}
where $V_{\rm eq} =\Omega R = 92.3$ m~s$^{-1}$ is the equatorial rotation speed of Ceres. The solar spectrum reflected by the $k$-th surface element and observed on the Earth is shifted by a total radial velocity $v_{k} = v_{k \odot} + v_{k \oplus}$. 

To compute the RV variations, we consider a mean local spectral line with a Gaussian profile of central depth of 0.45 and full width at half maximum of  7035 m~s$^{-1}$ that provide a good approximation to the cross-correlation function of the true solar spectrum as given by HARPS pipeline. The local line profile is Doppler shifted by  the radial velocity $v_{k}$ and the intensity of the local continuum is assumed to be:
\begin{equation}
I_{k} = I_{0} a_{k} A_{k} \mu_{k\odot} \mu_{k\oplus} U(\mu_{k\oplus}) U(\mu_{k\odot}),
\label{flux_eq}
\end{equation}
where $I_{0}$ is a constant giving the intensity in the case of normal illumination and reflection, $a_{k}$ the albedo of the surface element, $A_{k}$ its area, and  $U$  a function that is equal to 1 when its argument is positive and is zero otherwise, so that the local continuum is zero when the given surface element is not visible or  is not illuminated by the Sun. Eq.~(\ref{flux_eq}) assumes that the solar intensity reflected towards the observer is proportional to $\mu_{k\odot} \mu_{k\oplus}$ according to Lambert's law, but it does not include any opposition surge effect. To compute the mean line profile integrated over the illuminated portion of Ceres' disc, we sum up all the local line profiles by weighting them proportionally to the intensity of their local continuum. Then the RV is obtained by fitting a Gaussian to the mean integrated line profile. 
 
The synthetic RV values must be scaled to reproduce the amplitude of the observed RV modulation because of the assumptions of our model. 
%The scaling constant and the offset are computed by assuming that the RV varies between 109 and 124 m~s$^{-1}$. 
 { \bf The resulting model  is overplotted to the radial velocity measurements once corrected by the    offset of 115.4~m~s$^{-1}$ in  Fig \ref{fig1.pdf}. }As discussed previously,    this is the expected offset    and our model provides a fair reproduction of the data points once the amplitude of the synthetic radial velocity variations   are properly scaled.   
%The contributions of the individual maxima of Chamberlain et al.'s template are also shown separately in  Fig.  \ref{fig1.pdf}. 
Note that a spot produces no change in radial velocity at the phase when the sub-observer longitude is equal to its longitude    because this is when it  passes exactly through the central meridian of Ceres' disc and in fact in correspondence of the Chamberlain et al' s maxima there are no shifts in radial velocities. We found also that     minima in the  photometric curve do not  produce significant features in radial velocities.

\section{Discussion}
Although a model based on the average albedo variations as derived by Chamberlain et al.    can provide a general description of the RV modulation, in particular  matching the  radial velocities at certain longitudes, an intrinsic variability is required to explain the difference between the radial velocities measured in the two epochs, specifically to account for the non reproducibility of the radial velocities  with phase.
One possibility is an intrinsic change of albedo of the bright features   corresponding to Occator in July and to the   photometric maximum  at phase  $0^{\circ}$ in August. The variability occurs when they are seen in the approaching hemisphere.  For instance,  on its rise on the night of July 31, Occator was in the approaching hemisphere and when it reached a distance of $\sim 45^{\circ}$ from the central meridian the most negative radial velocity was observed, in line with the predicted RV minimum produced by a bright spot there (cf. Eqs.~\ref{vr_eqs1} and~\ref{vr_eqs}).   Note that a similar  negative radial velocity was observed when Ceres  completed an entire rotation and Occator became visible again. After the passage across the central meridian  the RV curve became close to the synthetic one computed with the average albedo distribution,  which implies  no contribution  to the radial velocity curve by this feature.    
Thus, a strong reflectance of the bright spot in Occator when it rose on the visible hemisphere and its subsequent fast decrease could mimic the  observed radial velocity variation in July.
Nevertheless,  Occator did not play any major role during the August observations implying a significant reduction in the albedo of this spot at that time. 
 To explain the variability in  August  a change   of the stronger  maximum of Chamberlain et al.'s  light curve  around longitude $0^{\circ}$   when  seen in the approaching hemisphere   is needed. This  produces  slightly more negative velocities  at  phases   $100^{\circ}$ to $0^{\circ}$.
%They seem to have  a high albedo when they are frozen, which  decreases significantly as they  melt  following the release of internal heat,  or simply  under solar irradiation.  }
  %{\bf It must be noted that the Occator bright spot does not coincide with the Chamberlain et al. photometric maxima,  but quite surprisingly its  
%longitude   $240^{\circ}$  corresponds to a minimum of  the light curve}. 
% Some additional bright features, also characterized by a rapid fading, should be added to reproduce the details of the observed RV variation on that night, specifically the persistent negative RV after Occator passage across the central meridian.

\citet{per15}   found  variations  in the slope of visible spectra at the level of    
2-3 percent over  1000 Angstrom with a variation in the relative reflectivity of  about  10 percent in the region between   500 and  800 nm.
Herschel detected water vapor plumes erupting off the surface of Ceres, which may come from volcano-like ice geysers  \citep{kup14}.   
The recent Dawn  observations  suggest  that the bright spots  could provide 
some atmosphere in this particular region of Ceres    confirming  Herschel's  water vapor detection \citep{wit15}.   
It has been noted  that    the spots  appear bright at dawn on Ceres while they seem to fade by dusk. That could mean that sunlight plays an important role,  for instance  by heating up ice just beneath the surface and causing it to blast off some kind of plume or other feature. 
% Dawn's principal investigator C.  Russell  reported that  {\it the haze comes back in a regular pattern, it  covers about half of the crater and stops at the rim.}  

 It is  possible  to speculate that  a volatile substance  could   evaporate  from the inside and  freezes when it reaches the surface in shade. When it  arrives on the illuminated hemisphere, the
patches may change quickly under the action of the solar radiation loosing most of its reflectivity power when it is in the receding hemisphere. This could explain why    we do not  see  an  increase in positive radial velocities,  but all the changes  in  the radial velocity curves are characterized by negative values.  After being melted by the solar heat, the 
patches can form again  during the  four-hours-and-a-half duration of the night, but not
exactly in the previous fashion, thus the RV curve varies from one rotation to the other. It is possible that the  cycle of evaporation and freezing could  last more than one rotational period and so the changes in the albedo which are responsible of the variations in the radial velocity. 
The photometry variability induced by  these patches  could have been below the detectability threshold of Chamberlain et al. of  $\sim 0.005$~mag but should be  noted  by Dawn. %One would expect variations of the order of 0.10 mag  in the V passband, if those features were stable and the remaining part of the asteroid had a uniform albedo. However, since these bright features are localized in the approaching part of the disc, they  are somewhat balanced by the remaining dark surface pattern,  thus reducing significantly the amplitude of the disc-integrated photometric variations. 
 Dawn   is able  to resolve the disk of 
Ceres and should be able to see the patches of ice as changes in the 
localized reflectivity. Indeed, after submission of this paper  \citet{2015Natur.528..237N}  reported the presence of localized bright areas  which are consistent with hydrate magnesium sulfates. They found a bright pit  on the floor of crater Occator that shows probable sublimation of water ice, producing haze clouds inside which  appear and disappear with a diurnal rhythm. In particular,  their figure 4  shows a diffuse haze that fills the floor of Occator and that disappears almost completely at dusk, and this we believe could be a physical framework  for  the daily variability we have detected in radial velocities. If a closer connection can be established between the two effects  it will be  an opportunity to monitor the daily activity in Ceres that can continue also beyond the space mission.

%and indeed people working on the 
%image analysis of Dawn pictures should have already found some
%variations of the surface features over time or, especially, with
%local insolation. 

In conclusion,  the observed  RV pattern  is likely associated with a change in the albedo of  the material producing the photometric light curve variations but with some additional contributions of  
more  contrasted features such as  the spots in Occator.   
With the observations at our disposal we   suggest   the presence of  short-term variability of Ceres' albedo on  timescale of days or months.
However,  this   has  testable  predictions and can be   confirmed and further refined  by means of dedicated radial velocity observations and a detailed analysis of Dawn's images. 

 \bibliography{ceres_biblio}

 \section*{Acknowledgments}

This program used  Director Discretionary Time at
ESO  and we warmly acknowledge  ESO director  for this opportunity together with the La Silla staff for the collaboration and  competence during the execution of the observations.  Discussions with   Ivo Saviane  and  Francesco Pepe at different stages of this work are also
acknowledged.

 \begin{table*}
\caption{Online table: Journal of July observations. 1$^{st}$ column is the time of the start of the exposure in UT. 2$^{nd}$ column is  the time in MJD. 3$^{rd}$ column is the radial velocity computed by the HARPS  pipeline. 4$^{th}$ column is the radial velocity corrected from the instrumental drift. 5$^{th}$  column is the BERV  of the HARPS pipeline. 6$^{th}$  column is the exposure time in seconds. 7$^{th}$  column is the flux weighted mid exposure fraction. 8$^{th}$  column is the effective mid exposure time. 9$^{th}$  column is the radial velocity of Ceres with respect to the Sun. 10$^{th}$  is  the radial velocity of the observer relative to Ceres.}             
\label{table:1}      
\centering          
\begin{tabular}{c c c  cc    cc   c c c  }     % 10 columns 
\hline\hline       
Date & MJD  & RV &   RV$_c$ & BERV  & Exp & T'  & JDMIDEXP & $\dot r$ & $\dot \delta$\\ 
 &    & m~s$^{-1}$  &    m~s$^{-1}$  &  m~s$^{-1}$    & s &   &  m~s$^{-1}$   &  m~s$^{-1}$   &   m~s$^{-1}$  \\ 
\hline  
2015-07-30 23:33:06 &  2457233.98138201 & 914.885 &	 914.578    &  -3203.246    &	   179.9  &	 0.50  &   2457234.48131   & 671.421  & 3357.268 \\
2015-07-30 23:51:29 &  2457233.99414785 & 917.027 &	 916.741    &  -3224.302    &	   779.9  &	 0.54  &   2457234.49408   & 671.371  & 3371.657 \\
2015-07-31 00:11:23 &  2457234.00353524 & 926.589 &	 926.437    &  -3236.302    &	   779.9  &	 0.49  &   2457234.50790   & 671.317  & 3389.635 \\
2015-07-31 00:25:32 &  2457234.01292216 & 927.485 &	 927.276    &  -3250.686    &	   779.9  &	 0.54  &   2457234.51773   & 671.279  & 3403.891 \\
2015-07-31 00:38:32 &  2457234.02230851 & 931.640 &	 931.080    &  -3264.919    &	   779.9  &	 0.50  &   2457234.52675   & 671.244  & 3418.031 \\
2015-07-31 00:52:08 &  2457234.03175053 & 932.855 &	 932.761    &  -3280.839    &	   779.9  &	 0.50  &   2457234.53620   & 671.207  & 3433.858 \\
2015-07-31 01:05:47 &  2457234.04113954 & 932.896 &	 932.570    &  -3297.837    &	   779.9 &	 0.51 &    2457234.54568  &  671.170 &  3450.765 \\
2015-07-31 01:19:18 &  2457234.05052624 & 934.205 &	 933.700    &  -3315.651    &	   779.9 &	 0.51 &    2457234.55506  &  671.134 &  3468.473 \\
2015-07-31 01:32:50 &  2457234.05991305 & 932.692 &	 932.504    &  -3334.361    &	   779.9 &	 0.51 &    2457234.56446  &  671.097 &  3487.118 \\
2015-07-31 02:02:21 &  2457234.08032856 & 924.051 &	 923.737    &  -3387.639    &	   779.9 &	 0.52 &    2457234.58496  &  671.018 &  3530.718 \\
2015-07-31 02:15:37 &  2457234.08971583 & 923.760 &	 923.602    &  -3408.513    &	   779.9 &	 0.50 &    2457234.59417  &  670.982 &  3551.508 \\
2015-07-31 02:27:26 &  2457234.09910253 & 921.673 &	 921.613    &  -3429.829    &	   779.9 &	 0.47 &    2457234.60238  &  670.950 &  3570.585 \\
2015-07-31 02:42:39 &  2457234.10848969 & 921.452 &	 921.234    &  -3453.032    &	   779.9 &	 0.50 &    2457234.61295  &  670.909 &  3595.859 \\
2015-07-31 02:57:52 &  2457234.11915158 & 922.022 &	 921.713    &  -3479.130    &	   779.9 &	 0.49 &    2457234.62351  &  670.868 &  3621.848 \\
2015-07-31 03:11:23 &  2457234.12853526 & 923.979 &	 923.878    &  -3502.820    &	   779.9 &	 0.49 &    2457234.63290  &  670.831 &  3645.457 \\
2015-07-31 03:24:45 &  2457234.13792207 & 922.025 &	 922.003    &  -3526.693    &	   779.9 &	 0.48 &    2457234.64218  &  670.795 &  3669.219 \\
2015-07-31 03:39:11 &  2457234.14730844 & 922.650 &	 922.747    &  -3552.847    &	   779.9 &	 0.55 &    2457234.65221  &  670.756 &  3695.265  \\
2015-07-31 04:05:28 &  2457234.16636159 & 926.459 &	 926.326    &  -3601.166    &	   779.9 &	 0.46 &    2457234.67046  &  670.685 &  3743.455  \\
2015-07-31 04:19:22 &  2457234.17574598 & 923.333 &	 923.457    &  -3626.983    &	   779.9 &	 0.49 &    2457234.68011  &  670.648 &  3769.190  \\
2015-07-31 04:33:08 &  2457234.18512527 & 928.590 &	 928.800    &  -3652.623    &	   779.9 &	 0.51 &    2457234.68967  &  670.611 &  3794.748  \\
2015-07-31 04:52:32 &  2457234.19904883 & 916.802 &	 916.704    &  -3702.054    &	   779.9 &	 0.46 &    2457234.70314  &  670.558 &  3830.735  \\
2015-07-31 05:07:29 &  2457234.20843368 & 916.405 &	 916.379    &  -3729.726    &	   779.9 &	 0.57 &    2457234.71353  &  670.518 &  3858.322  \\
2015-07-31 05:19:34 &  2457234.21781795 & 917.362 &	 917.358    &  -3751.936    &	   779.9 &	 0.46 &    2457234.72192  &  670.485 &  3880.455  \\
2015-07-31 05:33:36 &  2457234.22720429 & 918.551 &	 918.331    &  -3777.466    &	   779.9 &	 0.50 &    2457234.73166  &  670.448 &  3905.901  \\
2015-07-31 05:47:48 &  2457234.23680814 & 921.090 &	 921.056    &  -3802.994    &	   779.9 &	 0.53 &    2457234.74152  &  670.409 &  3931.290  \\
2015-07-31 06:00:49 &  2457234.24620873 & 917.738 &	 917.472    &  -3825.965    &	   779.9 &	 0.49 &    2457234.75056  &  670.374 &  3954.179  \\
2015-07-31 06:14:37 &  2457234.25559787 & 920.371 &	 920.174    &  -3849.816    &	   779.9 &	 0.51 &    2457234.76015  &  670.337 &  3977.972  \\
2015-07-31 06:27:51 &  2457234.26498376 & 919.552 &	 919.722    &  -3872.227    &	   779.9 &	 0.49 &    2457234.76934  &  670.301 &  4000.268  \\
2015-07-31 06:41:38 &  2457234.27445472 & 917.456 &	 917.350    &  -3894.926    &	   779.9 &	 0.50 &    2457234.77891  &  670.264 &  4022.884  \\
2015-07-31 06:55:41 &  2457234.28384802 & 916.585 &	 916.404    &  -3917.367    &	   779.9 &	 0.54 &    2457234.78866  &  670.226 &  4045.232  \\
2015-07-31 07:08:49 &  2457234.29323588 & 917.716 &	 917.553    &  -3937.646    &	   779.9 &	 0.51 &    2457234.79778  &  670.190 &  4065.418  \\
2015-07-31 07:21:48 &  2457234.30262281 & 918.190 &	 917.813    &  -3956.999    &	   779.9 &	 0.47 &    2457234.80680  &  670.155 &  4084.654  \\
2015-07-31 07:40:31 &  2457234.31507058 & 899.861 &	 899.612    &  -3996.341    &	   779.9 &	 0.53 &    2457234.81980  &  670.105 &  4111.025  \\
2015-07-31 07:54:10 &  2457234.32446365 & 902.315 &	 902.130    &  -4014.586    &	   779.9 &	 0.54 &    2457234.82928  &  670.068 &  4129.177  \\
2015-07-31 08:06:54 &  2457234.33384849 & 904.157 &	 903.830    &  -4030.755    &	   779.9 &	 0.48 &    2457234.83812  &  670.034 &  4145.240  \\
2015-07-31 08:20:32 &  2457234.34323427 & 900.641 &	 900.403    &  -4047.094    &	   779.9 &	 0.49 &    2457234.84759  &  669.997 &  4161.463  \\
2015-07-31 08:35:06 &  2457234.35335007 & 899.825 &	 899.475    &  -4063.372    &	   779.9 &	 0.49 &    2457234.85770  &  669.958 &  4177.637  \\
2015-07-31 08:48:30 &  2457234.36273677 & 898.333 &	 898.064    &  -4077.237    &	   779.9 &	 0.48 &    2457234.86701  &  669.921 &  4191.418  \\
2015-07-31 09:02:00 &  2457234.37212336 & 898.801 &	 898.521    &  -4090.155    &	   779.9 &	 0.48 &    2457234.87638  &  669.885 &  4204.205  \\
2015-07-31 09:14:37 &  2457234.38150995 & 894.008 &	 893.800    &  -4101.183    &	   779.9 &	 0.41 &    2457234.88515  &  669.851 &  4215.134  \\

 \hline      	    																						 
 \end{tabular}	    																						 
\end{table*}

 \begin{table*}
\caption{as Tab \ref{table:2}   for the August observations.}             
\label{table:2}      
\centering          
\begin{tabular}{c c c  cc    cc   c c c  }     % 10 columns 
\hline\hline       
Date & MJD  & RV  &   RV$_c$ & BERV  & Exp  & T'  & JDMIDEXP & $\dot r$ & $\dot \delta$\\ 
 &    &m~s$^{-1}$  &    m~s$^{-1}$  &  m~s$^{-1}$    & s &   &  m~s$^{-1}$   &  m~s$^{-1}$   &   m~s$^{-1}$  \\ 
 
\hline  
2015-08-26 23:36:05&  2457260.98339249 & -2056.98 &    -2057.768    &  -17583.323    &   179.99    &	0.50	&  2457261.48440    &564.365   & 14849.496  \\  
2015-08-26 23:42:28&  2457260.98783118 & -2058.01 &    -2058.517    &  -17602.255    &   899.99    &	0.54	&  2457261.49340    &564.329   & 14867.509  \\ 
2015-08-26 23:57:59&  2457260.99860058 & -2058.04 &    -2058.533    &  -17624.754    &   899.99    &	0.52	&  2457261.50400    &564.286   & 14889.613  \\  
2015-08-27 00:13:30&  2457261.00938260 & -2056.57 &    -2057.134    &  -17648.810    &   899.99    &	0.53	&  2457261.51490    &564.242   & 14913.262  \\  
2015-08-27 00:29:01&  2457261.02015688 & -2057.93 &    -2058.440    &  -17672.274    &   899.99    &	0.48	&  2457261.52510    &564.201   & 14936.154  \\  
2015-08-27 00:46:51&  2457261.03254109 & -2058.62 &    -2059.234    &  -17702.217    &   899.99    &	0.51	&  2457261.53780    &564.150   & 14965.563  \\  
2015-08-27 01:02:22&  2457261.04331652 & -2058.69 &    -2059.002    &  -17727.530    &   899.99    &	0.48	&  2457261.54830    &564.107   & 14990.529  \\  
2015-08-27 01:17:53&  2457261.05409346 & -2059.29 &    -2059.711    &  -17754.388    &   899.99    &	0.49	&  2457261.55920    &564.063   & 15016.965   \\ 
2015-08-27 01:33:24&  2457261.06486773 & -2059.02 &    -2059.422    &  -17781.380    &   899.99    &	0.49	&  2457261.56990    &564.020   & 15043.326   \\ 
2015-08-27 01:49:00&  2457261.07569454 & -2059.07 &    -2059.489    &  -17809.909    &   899.99    &	0.53	&  2457261.58120    &563.974   & 15071.493   \\ 
2015-08-27 02:04:30&  2457261.08646442 & -2059.27 &    -2059.706    &  -17836.587    &   899.99    &	0.50	&  2457261.59160    &563.932   & 15097.609   \\ 
2015-08-27 02:20:01&  2457261.09723998 & -2059.28 &    -2059.470    &  -17864.206    &   899.99    &	0.50	&  2457261.60240    &563.889   & 15124.820   \\ 
2015-08-27 02:35:32&  2457261.10801539 & -2059.29 &    -2059.802    &  -17891.783    &   899.99    &	0.50	&  2457261.61320    &563.845   & 15152.014   \\ 
2015-08-27 02:51:07&  2457261.11884118 & -2056.63 &    -2056.970    &  -17919.642    &   899.99    &	0.51	&  2457261.62410    &563.801   & 15179.332   \\ 
2015-08-27 03:06:38&  2457261.12961279 & -2055.70 &    -2055.898    &  -17947.668    &   899.99    &	0.54	&  2457261.63520    &563.756   & 15206.906   \\
2015-08-27 03:22:09&  2457261.14038834 & -2055.70 &    -2055.861    &  -17972.477    &   899.99    &	0.46	&  2457261.64510    &563.716   & 15231.195   \\
2015-08-27 03:37:40&  2457261.15116389 & -2056.25 &    -2056.495    &  -18000.224    &   899.99    &	0.51	&  2457261.65640    &563.671   & 15258.461   \\ 
2015-08-27 03:53:50&  2457261.16238518 & -2058.62 &    -2059.007    &  -18027.431    &   899.99    &	0.52	&  2457261.66780    &563.625   & 15285.359   \\ 
2015-08-27 04:09:20&  2457261.17315436 & -2056.75 &    -2057.070    &  -18052.635    &   899.99    &	0.52	&  2457261.67850    &563.581   & 15309.942   \\ 
2015-08-27 04:24:51&  2457261.18393002 & -2056.86 &    -2057.360    &  -18076.632    &   899.99    &	0.50	&  2457261.68910    &563.538   & 15333.567   \\ 
2015-08-27 04:40:22&  2457261.19470522 & -2057.71 &    -2057.788    &  -18100.291    &   899.99    &	0.50	&  2457261.69990    &563.495   & 15356.801   \\ 
2015-08-27 04:56:42&  2457261.20604863 & -2057.93 &    -2058.313    &  -18124.398    &   899.99    &	0.51	&  2457261.71130    &563.449   & 15380.308   \\ 
2015-08-27 05:12:13&  2457261.21682348 & -2057.41 &    -2057.577    &  -18145.619    &   899.99    &	0.49	&  2457261.72190    &563.406   & 15401.141   \\ 
2015-08-27 05:27:44&  2457261.22759891 & -2056.97 &    -2057.160    &  -18166.765    &   899.99    &	0.52	&  2457261.73300    &563.361   & 15421.814   \\ 
2015-08-27 05:43:16&  2457261.23838592 & -2058.20 &    -2058.406    &  -18185.583    &   899.99    &	0.49	&  2457261.74340    &563.319   & 15440.050   \\ 
2015-08-27 06:02:28&  2457261.25171811 & -2059.08 &    -2059.487    &  -18207.505    &   899.99    &	0.47	&  2457261.75660    &563.266   & 15461.516   \\ 
2015-08-27 06:17:59&  2457261.26249690 & -2062.50 &    -2062.708    &  -18224.963    &   899.99    &	0.53	&  2457261.76800    &563.220   & 15478.459   \\ 
2015-08-27 06:33:30&  2457261.27327013 & -2059.84 &    -2059.847    &  -18239.478    &   899.99    &	0.49	&  2457261.77830    &563.178   & 15492.437   \\
2015-08-27 06:49:01&  2457261.28404742 & -2061.54 &    -2061.933    &  -18253.307    &   899.99    &	0.50	&  2457261.78920    &563.134   & 15505.802   \\
2015-08-28 06:05:08&  2457262.25357302 & -2164.08 &    -2164.023    &  -18648.104    &   179.99    &	0.51	&  2457262.75860    &559.216   & 15808.005   \\
2015-08-28 06:13:37&  2457262.25946192 & -2169.41 &    -2169.494    &  -18663.502    &   899.99    &	0.50	&  2457262.76460    &559.192   & 15816.716   \\
2015-08-28 06:29:08&  2457262.27023769 & -2170.04 &    -2170.191    &  -18678.809    &   899.99    &	0.52	&  2457262.77560    &559.147   & 15831.567   \\
2015-08-28 06:44:39&  2457262.28101416 & -2171.18 &    -2171.264    &  -18691.769    &   899.99    &	0.47	&  2457262.78590    &559.105   & 15844.119   \\
2015-08-28 07:00:10&  2457262.29179040 & -2173.17 &    -2173.340    &  -18704.048    &   899.99    &	0.48	&  2457262.79670    &559.062   & 15855.832   \\
 \hline      	    																						 
 \end{tabular}	    																						 
\end{table*}

\end{document}